\documentstyle[12pt]{article}

\newskip\humongous \humongous=0pt plus 1000pt minus 1000pt
\def\caja{\mathsurround=0pt}

\newif\ifdtup
\def\panorama{\global\dtuptrue \openup1\jot \caja
	\everycr{\noalign{\ifdtup \global\dtupfalse
	\vskip-\lineskiplimit \vskip\normallineskiplimit
	\else \penalty\interdisplaylinepenalty \fi}}}
\def\eqalignno#1{\panorama \tabskip=\humongous
	\halign to\displaywidth{\hfil$\displaystyle{##}$
	\tabskip=0pt&$\displaystyle{{}##}$\hfil
	\tabskip=\humongous&\llap{$##$}\tabskip=0pt
	\crcr#1\crcr}}

\let\vev\VEV

\def\abs#1{\left| #1\right|}

\def\ltap{\raisebox{-.4ex}{\rlap{$\sim$}} \raisebox{.4ex}{$<$}}
\def\gtap{\raisebox{-.4ex}{\rlap{$\sim$}} \raisebox{.4ex}{$>$}}

\begin{document}
\begin{titlepage}
\begin{center}
\today          \hfill 
\hfill    LBL-38381 \\
          \hfill    UCB-PTH-96/07 \\

\vskip .5in

{\large \bf A Grand Unified Supersymmetric Theory of Flavor}
\footnote{This work was supported in part by the Director, Office of 
Energy Research, Office of High Energy and Nuclear Physics, Division of 
High Energy Physics of the U.S. Department of Energy under Contract 
DE-AC03-76SF00098 and in part by the National Science Foundation under 
grant PHY-95-14797.}

\vskip .25in

Riccardo Barbieri

{\em Physics Department, University of Pisa \\
    and INFN Sez. di Pisa, I-56126 Pisa, Italy}
\vskip .25in

Lawrence J. Hall

{\em Physics Department and Lawrence Berkeley Laboratory\\
      University of California, Berkeley, CA 94720}
\end{center}

\vskip .5in

\begin{abstract}
A grand unified $SU(5)$ theory is constructed with a hierarchical
breaking of a $U(2)$ flavor symmetry. The small parameters of the squark and 
slepton mass matrices, necessary to solve the supersymmetric flavor-changing 
problem, and the inter-generational quark and lepton mass hierarchies are
both generated from the $U(2)$ symmetry breaking parameters. The flavor
interactions of the theory are tightly constrained, with just 10 free real 
parameters for both the fermion and scalar sectors. All but one of the 8 small
fermion mass ratios, and all of the 3 small Cabibbo-Kobayashi-Maskawa mixing
angles, can be understood without introducing small dimensionless Yukawa
parameters. Predictions are made for 2 of the Cabibbo-Kobayashi-Maskawa mixing
angles and for 2 of the fermion masses. The six flavor mixing matrices which
appear at the neutralino vertices, and which in general are arbitrary unitary
matrices, are determined in terms of just a single free parameter.
\end{abstract}
\end{titlepage}
\renewcommand{\thepage}{\roman{page}}
\setcounter{page}{2}
\mbox{ }

\vskip 1in

\begin{center}
{\bf Disclaimer}
\end{center}

\vskip .2in

\begin{scriptsize}
\begin{quotation}
This document was prepared as an account of work sponsored by the United
States Government. While this document is believed to contain correct 
 information, neither the United States Government nor any agency
thereof, nor The Regents of the University of California, nor any of their
employees, makes any warranty, express or implied, or assumes any legal
liability or responsibility for the accuracy, completeness, or usefulness
of any information, apparatus, product, or process disclosed, or represents
that its use would not infringe privately owned rights.  Reference herein
to any specific commercial products process, or service by its trade name,
trademark, manufacturer, or otherwise, does not necessarily constitute or
imply its endorsement, recommendation, or favoring by the United States
Government or any agency thereof, or The Regents of the University of
California.  The views and opinions of authors expressed herein do not
necessarily state or reflect those of the United States Government or any
agency thereof, or The Regents of the University of California.
\end{quotation}
\end{scriptsize}

\vskip 2in

\begin{center}
\begin{small}
{\it Lawrence Berkeley Laboratory is an equal opportunity employer.}
\end{small}
\end{center}

\newpage
\renewcommand{\thepage}{\arabic{page}}
\setcounter{page}{1}
\noindent {\bf 1. The flavor group U(2)}

The fermion mass puzzle arose with the discovery of the muon, and has become
more pressing with the discovery of each new quark and lepton.
In terms of the standard model, the question is: what is the origin of the small
dimensionless parameters in the Yukawa coupling matrices?
In supersymmetric extensions of the standard model, the spectrum of squarks and
sleptons possess a second puzzle.
Although none of these particles have masses much less than the weak scale, the
scalar mass matrices are highly constrained by flavor-changing processes
\cite{DG}, and must involve a second set of small dimensionless parameters.

The fermion and scalar mass matrices are different aspects of the supersymmetric
flavor problem, so that it is attractive to consider these two sets of
small parameters to be related. 
The key to such a relationship is provided by flavor symmetries.

A flavor group $G_f$, which commutes with supersymmetry, treats quarks and
squarks identically.
In the $G_f$ symmetric limit the squarks acquire masses, but have mass matrices
with a high degree of flavor conservation, while the quarks are massless, except
possibly the heaviest ones.
The lighter quark masses are generated when $G_f$ is broken hierarchically by a
set of vevs, $v_i$, so that the small parameters of the Yukawa matrices involve
$v_i/M_f \equiv \epsilon_i$, where $M_f$ is a flavor mass scale.
Such breakings also introduce corrections to the squark mass matrices, some of
which violate flavor.
However, these flavor-changing effects are proportional to $\epsilon_i$, and are
suppressed for the same reason that some quarks are light.
Such a mechanism deserves the title ``super-GIM'' \cite{GIM}.

The power and simplicity of this use of approximate flavor symmetries was first
illustrated using $G_f = U(3)^5$,
the maximal flavor group of the standard model, with the $\epsilon_i$ taken to
be the three Yukawa matrices \cite{HRa}.
Such a scheme, called effective weak scale supersymmetry, provides a framework
for the soft operators which is greatly preferable to the
universality assumption.
However, this scheme treated the Yukawa matrices as phenomenological symmetry
breaking parameters, and did not provide a theory for their origin.
Several such models have been constructed over the last three years
[4-15],
based on flavor groups which are Abelian or non-Abelian,
continuous or discrete, and gauged or global.

We consider this development $-$ the ability to construct supersymmetric 
theories
of flavor $-$ to be of great importance. For quark and lepton masses it provides
a symmetry basis for textures, which need no longer be postulated purely on
grounds of phenomenological simplicity. Not only can these theories solve the
flavor-changing problem, but the coupling to the fermion mass problem produces
a very constrained framework. In the present paper, we continue our attempt to
develop a theory with a simple believable symmetry structure, which solves the
flavor-changing problem, provides an economical description of 
the quark and lepton spectrum, and is able to make experimentally testable
predictions, both in the fermion and scalar sectors.

Three requirements provide a guide in choosing the flavor group, $G_f$. 

\begin{enumerate}
\item $G_f$ {\it must solve the flavor-changing problem.}

The minimal, most straightforward and compelling flavor symmetry solution 
to the flavor-changing problem is for $G_f$ to be
non-Abelian, with the lightest two generations in doublets
$$
q_a = \pmatrix{q_1\cr q_2} \; \; u_a = \pmatrix{u_1\cr u_2} \; \; d_a = 
\pmatrix{d_1\cr d_2} \;
\; \ell_a =\pmatrix{\ell_1\cr \ell_2} \; \; e_a = \pmatrix{e_1\cr e_2}.\eqno(1)
$$
If this symmetry is sufficiently weakly broken, the resulting near degeneracy of
the scalars solves the flavor-changing puzzle.
\footnote{Flavor changing amplitudes are also induced by a non-degeneracy
between the scalars of the third generation and those of the lighter two
generations. These effects, although close to the limits of what experiments
allow, are not problematic if the relevant mixing angles are similar to the
corresponding CKM mixings and/or the amount of fractional mass splitting is 
somewhat less than maximal.}
We find it surprising that this elegant idea was not studied prior to 1993, when
$G_f = SU(2)$ was considered \cite{DLK}.

\item $G_f$ {\it must be compatible with gauge unification.}

There are many groups which could have the representation  structure of (1).
The choice can be greatly reduced by requiring that the group acts identically 
on $(q_a, u_a, d_a, \ell_a, e_a)\equiv \psi_a$, as results from a theory in 
which the components of a generation are unified.

\item {\it In the symmetric limit, fermions of the first two generations must
be massless.}

The flavor group $G_f = SU(2)$ allows the interaction $q_a\epsilon^{ab}
d_b h$, giving unacceptable, large, degenerate masses to $d$ and $s$ quarks.
We are therefore led to consider $G_f = U(2)$, which can be written as $SU(2)
\times U(1)$ with $\psi_a$ transforming as (2,1).
The tensor $\epsilon^{ab}$ is a non-trivial singlet of $U(2)$ carrying charge
-2, so that $U(2)$ invariance allows Yukawa couplings only for the third
generation, which is taken to transform as a trivial $U(2)$ singlet.
\end{enumerate}

A discrete subgroup of $U(2)$ might provide an acceptable alternative choice for
$G_f$. We prefer the 
continuous groups, however, because $U(2)$ contains a $U(1)$
subgroup with a color anomaly.
The Peccei-Quinn solution to the strong CP problem \cite{PQ} arises as an
automatic consequence of the above three requirements, which led us to choose
$G_f = U(2)$.
The strong CP problem involves the phase of the determinant of the quark mass
matrix, and hence is clearly an aspect of the flavor problem.
The Peccei-Quinn symmetry naturally finds a home as a subgroup of a more
comprehensive flavor group.
This solution of the strong CP problem would be lost if $U(2)$ were gauged.
Gauging a continuous flavor group is problematic, 
however, as  the $D^2$ contribution to the scalar masses 
reintroduces the flavor-changing problem \cite{M}.
We are therefore led to a non-Abelian, continuous, global flavor group: $G_f
=U(2)$.

While we believe the choice of $G_f = U(2)$ is very  well motivated, it is
obviously not unique.
For example, $U(2)$ could be extended to $U(3)$, with the three generations
forming a triple  $(\psi_a, \psi_3)$.
We view $U(2)$ as a stage of partial flavor unification.
We prefer to study $U(2)$ first: the top quark mass strongly breaks $U(3)$ to 
$U(2)$, and hence it is the weakly broken $U(2)$ which must solve the 
flavor-changing and fermion mass hierarchy problems.
It is important to establish whether $U(2)$ theories can solve these
problems.
While the representation structure (1) appears promising, a general low energy
effective $U(2)$ theory does not solve the flavor changing problem \cite{PT}.

A complete $U(3)$ flavor-unified theory would not only be elegant, but it also
offers the  prospect of a flavor symmetry origin for $R$ parity, which
$U(2)$ alone is unable to provide, since matter parity is a parity of $U(3)$ 
triality \cite{CHM}.

Although Abelian symmetries can constrain the mass matrices to solve the
flavor-changing problem \cite{NS}, we find the necessary group structure to be
less compelling than that of $U(2)$ or $U(3)$, due to a large freedom in the
choice of charge quantum numbers. For example, the rank 2 case of  $G_f = U(1)
^2$ contains two symmetry breaking parameters, $\epsilon_1$ and $\epsilon_2$,
which can appear in a mass matrix element as $\epsilon^n_1\epsilon^m_2$,
where $n$ and $m$ are positive integers which can be freely chosen by suitable
charge assignments.
Compare this to the rank 2, non-Abelian care of $G_f=U(2)$, which also has 2
symmetry breaking parameters, $\epsilon$ and $\epsilon'$, which we find appear
only linearly in the Yukawa matrices. Indeed, while the small parameters
$\epsilon $ and $\epsilon'$ solve the flavor-changing problem and account for
the two intergenerational fermion mass hierarchies, they are unable to describe
all the features of the quark and lepton mass matrices. 
Nevertheless, we find that the highly constrained group theory, and the
resulting testable predictions, are an important virtue of the $U(2)$ theory.
In this paper we seek to understand several other features of the quark and
lepton mass matrices from the SU(5) unified gauge symmetry.

\medskip

\noindent{\bf 2. The Structure of $U(2)$ Theories.}

\medskip

In the next sections we discuss in detail
the simplest $U(2)$ models and their predictions.
In this section we discuss general aspects of the construction of models with
$G_f = U(2)$.

The generations are assigned to $\psi_a(2) + \psi_3(1)$, where $\psi$ 
represents $q,u,d,\ell, $ or $e$, and does not imply any particular choice of 
gauge group.
We choose the two light Higgs doublets, $h$, to be $G_f$ singlets, both for 
simplicity and because, in the $U(3)$ extension of the flavor group, this 
allows for a flavor symmetry origin of matter parity.
The renormalizable superpotential  contains Yukawa couplings only for the third
generation, $\psi_3 \psi_3 h$, and the first question is therefore how $U(2)$ 
breaking can lead to a 23 entry for the Yukawa matrices.

The only known way of generating small dimensionless parameters is from
perturbative loop factors or from ratios of mass scales. A radiative origin for
$m_e/m_\mu$ in a theory with $G_f = U(2)$ has been discussed elsewhere
\cite{ACH}, in this paper we consider the fermion hierarchies to arise from 
a set of flavon vevs which break the flavor group at scales beneath some 
flavor scale $M_f$. From the viewpoint of an effective theory beneath $M_f$, 
it is clear that the 23 entry of the Yukawa matrices must come from 
an interaction of the form
$$
{1 \over M_f} \; [\psi_3\phi^a\psi_a h]_F \eqno(2)
$$ 
where $\phi^a$ is a doublet flavon, with opposite $U(1)$
charge to $\psi_a$, taking a vev $\vev{\phi^a} = (O, V)$.
The most general effective theory would also contain interactions quadratic in
$\phi^a$:
$$
{1 \over M_f^2} \; [\psi_a\phi^a\phi^b\psi_b \; h]_F \eqno(3)
$$
and
$$
{1 \over M_f^2} \;[\psi^{\dagger a} \phi^\dagger_a \phi^b \psi_b \; 
z^\dagger z]_D,\eqno(4)
$$
where $z$ is a supersymmetry breaking spurion, taken dimensionless,
$z=m \theta^2$.
Operators (2) and (3) lead to masses for second generation fermions at order 
$\epsilon^2$,
where $\epsilon = V/M$, while (4) leads to a non-degeneracy between the 
scalars of the first two generations which is also of order $\epsilon^2$.
Hence in the lepton sector
$$
{m^2_{\tilde{e}} - m^2_{\tilde{\mu}}\over m_{\tilde{e}}^2 + m^2_{\tilde{\mu}}}
\approx O\pmatrix{m_\mu\over m_\tau}\eqno(5)
$$
and in the down quark sector
$$
{m^2_{\tilde{d}} - m^2_{\tilde{s}}\over m^2_{\tilde{d}} + m^2_{\tilde{s}}}
\approx O \pmatrix{m_s\over m_b}.\eqno(6)
$$
When combined with rotations in the 1/2 space to diagonalize the fermion 
mass matrix, these non-degeneracies are extremely problematic for $\mu \to 
e\gamma$ and $\epsilon_K$ \cite{PT}.

The general effective field theory based on $G_f = U(2)$ 
leads to difficulties.
However, an important point with regard to constructing supersymmetric theories 
of flavor is that {\it specific models, especially if they are simple, 
typically do not lead to the most general set of} $G_f$ {\it invariant 
operators of the low energy effective theory.}
This result has been crucial in several models which have been constructed 
\cite{LNS2,ADHRS,CHM,BDH}.
For supersymmetric theories of flavor, low energy effective theories are useful
only if they can be used to demonstrate that certain symmetry schemes are safe
from flavor-changing problems.
If a general effective theory has problematic flavor-changing properties, 
it simply tells us which operators should be avoided in constructing
explicit models.

In this paper we generate small Yukawa couplings, from $\vev{\phi}/M_f$, by
rotating from flavor to mass eigenstates \cite{FN}.
Let $\psi$ represent the light matter of $q,u,d,\ell,e$, where
for now we omit flavor indices.
Suppose that it has a Yukawa coupling $\chi h \psi$ to a Higgs doublet $h$ and 
some heavy matter $\chi = Q,U,D,L,E$. 
The heavy generations are vector-like, with mass terms $M \bar{\chi}\chi$.
Finally, mass mixing between light and heavy matter is induced by 
$\vev{\phi} = \epsilon M_f$ via the interaction $\bar{\chi}\phi\psi$ 
(as always in this paper,
we assume that the flavons, $\phi$, are gauge singlets).
The theory is described by the superpotential
$$
W = \bar{\chi}( M_f \chi+\phi\psi) + \chi h \psi\eqno(7)
$$
where coupling constants of order unity are understood.\footnote{Since 
$\phi$ is non-trivial under $G_f$, $\chi$ and $\psi$ are
typically distinguished by $G_f$. In the cases where they have the same $G_f$
transformation, $\chi$ is defined as the linear combination which has a bare
mass coupling to $\bar{\chi}$.}
The vev $\vev{\phi}$ implies that the heavy state is $\chi' = \chi + 
\epsilon \psi$ while the light 
matter is $\psi'=\psi-\epsilon \chi$ rather than $\psi$, so that when the heavy 
mass eigenstate is decoupled the interaction $\chi h \psi$ contains a small 
Yukawa coupling for $\psi': \epsilon \psi' h \psi'$.
The small parameter $\epsilon $ arises because $G_f$ is broken at a scale 
less than $M_f$.

This mass mixing of states introduces a similar non-trivial effect in the 
soft supersymmetry breaking interactions.
If $\psi$ and $\chi$ have different $G_f$ transformation properties the soft 
${\bf{m}}^2$
matrix is diagonal, with entries $m^2_\psi, m^2_\chi$.
Rotating from the flavor basis $(\psi, \chi)$ to the mass basis 
$(\psi', \chi')$, one finds:
$$\pmatrix{m^2_\psi &O\cr
O&m^2_\chi} \to \pmatrix{m^2_\psi+\epsilon^2m^2_\chi
&\epsilon(m^2_\chi-m^2_\psi)\cr
\epsilon(m^2_\chi - m^2_\psi) & m^2_\chi+\epsilon^2 m^2_\psi}.\eqno(8)
$$
On decoupling the heavy eigenstate $\chi'$, only the $m^2_\psi + 
\epsilon^2 m^2_\chi$ entry of this matrix is of interest.
When flavor indices are reintroduced, this entry is a $3 \times 3$ matrix,
and the $\epsilon^2 m^2_\chi$ terms can lead to non-degeneracies and 
flavor-changing entries at order $\epsilon^2$ \cite{DP}. If $\chi$ and $\psi$
have the same $G_f$ transformation, there are additional order $\epsilon$
contributions to the $\psi'^\dagger \psi'$ mass matrix, which arise from an
initial $\chi^\dagger \psi$ operator.

The generation of interactions involving light eigenstates, 
suppressed by powers of $\epsilon$, from interactions that involved the
initial $\chi$ flavor eigenstate, can be summarized by
$$
\eqalignno{
[\chi \psi h]_F &\to \epsilon [ \psi\psi h]_F&(9a)\cr
[\chi \psi h \; z]_F &\to \epsilon [\psi\psi h \; z]_F&(9b)\cr
[\chi^\dagger \chi \; z^\dagger z]_D &\to \epsilon^2  
[ \psi^\dagger \psi \; z^\dagger z]_D&(9c)\cr}
$$
where (9b) yields soft trilinear scalar interactions. 
An immediate consequence of this picture is that there are no scalar mass 
terms linear in $\epsilon$.
For example, the operator $[\psi^\dagger_3 \phi^a \psi_a]_D$ can never be 
generated by this mechanism. 

It is frequently useful to use an approximate diagrammatic technique to 
perform the generation of the operators
9a, 9b, 9c from diagonalization of heavy mass matrices.
This is especially true for models more complicated than the 
simplest example discussed here. 
The three diagrams for 9a, 9b and 9c are shown in Figures 1a, 1b, 1c.
If $\chi^2$ contains a $G_f$ singlet, additional $O(\epsilon^2)$ contributions
to the Yukawa couplings amongst the light states result from:
$$
[\chi \chi h]_F \to \epsilon^2 [ \psi\psi h]_F \eqno(10)
$$
as illustrated in Figure 2. Such $O(\epsilon^2)$ contributions to Yukawa 
matrices
are more dangerous than the  $O(\epsilon)$ contributions of (9a) from Figure
1a: to get a particular value for a Yukawa coupling, they require a larger
value of $\epsilon$ and hence the scalar mass operators of (9c) lead to larger 
flavor-changing effects.

In this paper we consider only ``first order" Froggatt-Nielsen mixing, as
described above. In this case the mixing from a $\chi$ state, which has a
coupling to the Higgs, to an external $\psi$ state is linear in flavon fields.
Theories in which more powers of $\phi$ appear between Higgs and external
states are possible, by having a chain of internal heavy states of differing
$G_f$ quantum numbers. In this paper we do not consider theories with higher 
order
mixings: generally they are expected to be more dangerous than theories with
just first order mixing because the higher the order of the mixing the larger
the $\epsilon$ necessary to give the observed fermion masses.
 
We now consider the case of $U(2)$ where the external $\psi$ states are 
$\psi_a$ and $\psi_3$, and the Higgs field $h$ is a $U(2)$ singlet.
The 23 and 22 entries of the Yukawa coupling matrices cannot arise from the
diagram of Figure 2, because then the contributions of Figure 1c to the scalar
masses lead to the disastrous splittings of (5) and (6). This result is
independent of the U(2) representation choices for the $\chi$ and $\phi$
fields.

The 23 entry of the Yukawa matrices must be generated by Figure 1a, so that a
U(2) doublet flavon, $\phi^a$ is necessary and 
the operator in (9a) is $[\psi_3 \phi^a \psi_a h]_F$.
What are the $U(2)$ properties of $\chi$?
There are just two possibilities, either it is a singlet, $\chi$, or a 
doublet $\chi^a$.
The choice is critical, from the diagram of Fig. 1c it is immediately clear
that the singlet
$\chi$ exchange generates the dangerous operator (4), while the doublet $\chi^a$
exchange generates a 
harmless contribution to the third generation scalar mass: 
$[\psi^\dagger_3 \psi_3 \phi^\dagger_a \phi_a z^\dagger z]_D$.
A solution to the flavor-changing problem, based on the flavor group 
$U(2)$ alone, dictates that there should be no singlet $\chi$
states. Given the necessity of the doublet flavon, $\phi^a$, there can
similarly be no $\chi^a_b$ states.

A 22 entry for the Yukawa matrices can only be generated
from Figure 1a, which requires $(\phi, \chi) = (\phi^{ab}, \chi^a)$, where
$\phi^{ab}=+\phi^{ba}$, $\vev{\phi^{22}} \neq 0$. 
In this case the splitting in mass of
the scalars of the first two generations is quadratic in the
second generation fermion mass:
$$
{m^2_{\tilde{e}} - m^2_{\tilde{\mu}}\over m_{\tilde{e}}^2 + m^2_{\tilde{\mu}}}
\approx O\pmatrix{m_\mu^2 \over m_\tau^2}\eqno(11)
$$
and similarly for the up and down sectors. This gives contributions to $\mu
\rightarrow e \gamma$ and $\epsilon_K$ which are acceptable, although close to
the limit of what experiments allow. In this paper we construct the minimal
$U(2)$ model, in which there is no two index symmetric tensor $\phi^{ab}$.
   
Finally we consider generating Yukawa matrix elements which involve the
lightest generation. In principle these could originate from the diagram of
Figure 2, which involves $\chi$ states with zero $U(1)$ charge: $\chi, \chi^a_b,
\chi^{ac}_{bd}...$. However, the large vev of $\phi^a$, necessary for $V_{cb}$,
implies that $\chi$ and $\chi^a_b$ should be absent, so such diagrams would
necessarily involve $\chi$ states with at least four tensor indices, and
therefore $\phi$ states with at least three tensor indices. Ignoring such
complicated possibilities, all contributions to the Yukawa matrices arise from
Figure 1a, and therefore from the exchange of doublet $\chi$ states: $\chi^a$.
Hence, assuming no second order Froggatt-Nielsen mixing, the only question is
how many such $\chi^a$ states there are. Even this is only relevant in the case
of a unified gauge group where gauge breaking enters the masses of the $\chi^a$
states non-trivially. In this paper we consider a single $\chi^a$ state.

The most general contributions to Yukawa matrices from Figure 1a therefore 
involve
$(\phi; \chi) = (\phi^a, S^{ab}, A^{ab}; \chi^a)$ where $S^{ba} = + S^{ab}$
and $A^{ba} =- A^{ab}$.
The corresponding mixing of states is described by
$$
\overline{\chi}_a(M_f \chi^a + \phi^a \psi_3 + S^{ab} \psi_b + A^{ab} \psi_b).
\eqno(12)
$$
Allowing for the most general possible vevs of these flavons, this leads to 
Yukawa matrices of the form
$$
{\mbox{\boldmath$\lambda$}} = \pmatrix{ 0&0&0\cr
0&0&0\cr
0&0&1} + {1\over M_f} \pmatrix{ S^{11}&S_+&\phi^1\cr
S_-&S^{22}&\phi^2\cr
\phi^1&\phi^2&0}\eqno(13)
$$
and scalar mass matrices, from Figure 1c, of the form 
$$
{\bf m}^2 = 
\pmatrix{ m^2_1&0&0\cr
0&m^2_1&0\cr
0&0&m^2_3}\\
+ {1\over M^2_f}
\pmatrix{S^2_+ + (S^{11})^2 & S^{11} S_- + S^{22}S_+ & \phi^1 S^{22} + 
\phi^2S_+\cr
S^{11}S_- + S^{22}S_+ & S^2_- + (S^{22})^2 & \phi^2 S^{22} +\phi^1 S_-\cr
\phi^1 S^{11} + \phi^2 S_+ & \phi^2 S^{22} + \phi^1 S_- & (\phi^1)^2 + 
(\phi^2)^2} \eqno(14)
$$
where the fields stand for their vevs, and $S_{\pm} = S^{12} \pm A^{12}$.
The trilinear soft scalar interactions from Figure 1b take the form of (13).
The flavor-changing effects from this general scheme, which invokes only 
$\chi^a$ states, are acceptable: the exchange of scalars of the lighter two
generations give effects which are automatically well below expermental limits.
Flavor changing amplitudes are also induced by a non-degeneracy
between the scalars of the third generation and those of the lighter two
generations. These effects, although close to the limits of what experiments
allow, are not problematic if the relevant mixing angles are similar to the
corresponding CKM mixings and/or the amount of fractional mass splitting 
is somewhat less than maximal.

In this paper, rather than studying the most general doublet $\chi^a$ scheme 
given by (12), (13),
and (14), we study the very simplest such scheme, in which $S^{ab}$ is absent.
Several interesting phenomenological features follow from the vanishing of 
the 22 entry. \footnote{The case of $S^{22} \neq 0$ will be discussed
elsewhere.}
In this case, since $\vev{A^{12}}$ preserves $SU(2)$, $\vev{\phi^a}$ can 
be chosen to lie in the $a=2$ direction.
The Yukawa matrices and scalar mass matrices then depend on only two flavor 
vevs; \
$\epsilon = \vev{\phi^2}/M_f$ and $ \epsilon' = \vev{A^{12}}/M_f$, 
and take the forms
$$
{\mbox{\boldmath$\lambda$}} = \pmatrix{ 0&\epsilon'&0\cr
-\epsilon'&0&\epsilon\cr
0&\epsilon &1}\eqno(15)
$$
and 
$$
{\bf m}^2 = \pmatrix{ m^2_1 + \epsilon'^2 m^2 & 0 &\epsilon\epsilon' m^2\cr
0& m^2_1 + \epsilon'^2m^2 & 0\cr
\epsilon\epsilon' m^2 & 0 & m^2_3 + \epsilon^2 m^2}. \eqno(16)
$$
In (13) - (16) it is understood that each mass mixing entry involves 
an unknown $0(1)$ coefficient.
However, the $\epsilon'$ terms of (15) are antisymmetric, and
the two $\epsilon'^2m^2$ terms of (16) are identical since they 
do not violate $SU(2)$, \footnote{The coefficients of $S_+$ and $S_-$ 
of (13) are also equal, as are the coefficients of $S^2_+$ and $S^2_-$ 
of (14).} hence
$$
{m^2_{\tilde{e}} - m^2_{\tilde{\mu}}\over m_{\tilde{e}}^2 + m^2_{\tilde{\mu}}}
\approx O\pmatrix{m_e m_\mu^2 \over m_\tau^3}\eqno(17)
$$
A $U(2)$ flavor symmetry which solves the flavor-changing problem of 
supersymmetry provides a powerful tool for constraining the
flavor sector of supersymmetric theories.
Assuming only that the Higgs doublets are trivial under $U(2)$, and that 
more complicated higher
order mixings are irrelevant, we have shown that the entire flavor structure 
is generated from doublet $\chi^a$ exchange, as shown in (12), (13) and (14).
Furthermore, the assumption that $S^{ab}$ is absent leads to the remarkably 
simple theory of (15) and (16).
It is this theory that was introduced in \cite{BDH}, and in this paper we 
study further consequences of this theory in the case that the 
gauge group is grand unified.

\vskip 9pt

\noindent{\bf 3. The Minimal $U(2)$ Symmetric Model.}

\medskip

In this section we review the minimal $U(2)$ flavor structure in the case that
the gauge group is $SU(3) \times SU(2) \times U(1)$.
These results were obtained in reference \cite{BDH}.

The theory is defined by the interactions of (12) and (13), with the $S^{ab}$
tensor absent:
$$
\overline{\chi}_a (M_f\chi^a + \phi^a\psi_3 + A^{ab}\psi_b) + h(\psi_3\psi_3 +
\chi^a \psi_a).\eqno(18)
$$
Each of the matter fields ($\psi_3, \psi_a, \chi^a, \overline{\chi}_a$) contains
all components of a generations: $q, u^c, d^c, \ell, e^c$, or the conjugate
representations in the case of $\overline{\chi}_a$,
which we represent by
the index $i$, and $h$ represents both light Higgs doublets.
In (18), the coupling constants and their $i$ dependence
are left understood; hence
$\overline{\chi}_a\phi^a\psi_3 = \lambda_i \overline{\chi}_{ai}\phi^a \psi_{3i},
\; h\chi^a\psi_a = \lambda'_{ij} h \chi^a_i\psi_{aj} 
\; (ij= q u^c, u^c q, q d^c, d^cq, le^c, e^cl)$, etc.

The texture of the Yukawa and scalar trilinear matrices, 
$\mbox{\boldmath$\lambda$}$ and $\mbox{\boldmath$\xi$}$, 
is given in (16), and that of the scalar masses in (17).
The off-diagonal $\epsilon\epsilon'm^2$ entries of (17) 
are numerical insignificant, and can be dropped.
The diagonal correction terms, $\epsilon'^2 m^2$ and $\epsilon^2m^2$, can be
reabsorbed into the definition of the $m^2_1$ and $m^2_3$ parameters, so that
the scalar mass matrices are
$$
{\bf m}^2_i = \pmatrix{ m^2_1 &0&0\cr
                       0 & m^2_1 & 0\cr
                       0 & 0 &m^2_3} _i.\eqno(19)
$$
The scalars of the first two generations are accurately degenerate, and the
${\bf m}^2$ matrices involve 10 free parameters $m^2_{1i}$ and $m^2_{3i}$.

The Yukawa interactions are $\psi^T_I \mbox{\boldmath$\lambda$}_I \psi^c_I$, 
involving coupling matrices 
$$
{\mbox{\boldmath$\lambda$}}_I = \pmatrix{0&-D e^{i\phi_D}& 0\cr
                           D e^{i\phi_D} &0 & C e^{i\phi_C}\cr
                            0 & B e^{i\phi_B} & A e^{i\phi_A}}_I\eqno(20)
$$
where $I = U,D,E$ labels up, down and charged lepton sectors, and $A_I, B_I,
C_I$ and $D_I$ are real and positive.
The phases of these matrices can be factored into diagonal phase matrices ${\bf
P}$ and ${\bf P}^c$:
$$
{\mbox{\boldmath$\lambda$}}_I = {\bf P}_I \pmatrix{ 0 & -D & 0\cr
                                      D & 0 & C\cr
                                      0 & B & A}_I {\bf P}^c_I\eqno(21)
$$
where
$$
{\bf P}_I = \pmatrix{ e^{-i\alpha}& 0 & 0\cr
                      0& e^{i\beta} & 0\cr
                      0 & 0 & 1}_I\eqno(22a)
$$
and
$$
{\bf P}^c_I = \pmatrix{ e^{i(\phi_D-\beta)} & 0 & 0\cr
                        0 & e^{i\phi_B} & 0\cr
                        0 & 0 & e^{i\phi_A}}_I\eqno(22b)
$$
where
$\alpha_I = (\phi_B - \phi_D)_I$ and $\beta_I = (\phi_C - \phi_A)_I$.
Superfield phase rotations can remove all phases, except
$\alpha = \alpha_U - \alpha_D$ and $\beta =\beta_U - \beta_D$,
which appear only in charged current interactions.

The Yukawa matrices can be diagonalized by orthogonal rotations
$$
\pmatrix{0&-D&0\cr
         D&0&C\cr
         0&B&A}_I = {\bf R}_{23_I}{\bf R}_{12_I} 
\pmatrix{-{AD^2\over BC}&0 & 0\cr
         0 & -{BC\over A} & 0\cr
         0 & 0 &A}_I {\bf R}^{c^T}_{12_I} {\bf R}^{c^T}_{23_I}\eqno(23)
$$
so that the flavor mixing matrices, ${\bf W}_I$ and ${\bf W}^c_I$, appearing at
neutral gaugino $(\widetilde{\lambda})$ vertices, 
$
\widetilde{\psi}^\dagger_I {\bf{ W}}_I \psi_I \widetilde{\lambda}$ and
$\widetilde{\psi}^{c\dagger}_I {\bf{W}}^c_I \psi^c_I \widetilde{\lambda}$, 
are given, in the mass basis, by 
$$
\eqalignno{
{\bf{W}}^{(c)}_I &= {\bf{R}}^{(c)}_{23_I} {\bf{R}}^{(c)}_{12_I} = \pmatrix{
1&0&0\cr
0&1&s_{23}\cr
0&-s_{23} &1}^{(c)}_I\pmatrix{1&s_{12}&0\cr
                              -s_{12}& 1&0\cr
                              0&0&1}^{(c)}_I\cr
& = \pmatrix{1&s_{12}&0\cr
                                                        -s_{12} & 1 & s_{23}\cr
                                                        s_{12}s_{23} & -s_{23} &
                                                        1}^{(c)}_I. &(24)\cr}
$$
We have assumed that $(B/A)_I, (C/A)_I \approx 0(\epsilon )$ and
$(D/A)_I \approx 0(\epsilon')$ with $\epsilon' \ll \epsilon \ll 1$ so that the
small angle approximation is always valid.
We will find later that this is not necessarily always true.
The minimal $U(2)$ theory, in this approximation, has the interesting feature
that $W^{(c)}_{I_{13}} =0$.
Thus, for example, the photino vertex contains $\widetilde{\tau}^*e$ but not
$\widetilde{e}^*\tau$; staus can be made in electron collisions, but
selectrons will not decay to taus. 

The antisymmetry of the 12 entry of the Yukawa matrices implies 
$$
s^c_{12_I} = -s_{12_I}.\eqno(25)
$$
The angles of the mixing matrices arise from the diagonalization of the fermion
mass matrices, and depend on the fermion mass eigenvalues and the three free
parameters $r_I = (C/B)_I$:
$$
(s_{12})_I = \left( \sqrt{ {m_1\over m_2}}\right)_I\eqno(26a)
$$
$$
(s_{23})_I = \left( \sqrt{ r {m_2\over m_3}}\right)_I\eqno(26b)
$$
$$
(s^c_{23})_I = \left( \sqrt{ {1\over r} {m_2\over m_3}}\right)_I \eqno(26c)
$$
where $(m_{1,2,3})_I$ are the fermion mass eigenvalues of generations (1,2,3),
renormalized at the flavor scale $M_f$.
Choosing $A, B, C, D$ positive allows $\theta_{12}, \theta_{23}$ and 
$\theta_{23}^c$ to be taken in the first quadrant.

The trilinear scalar matrices, $\mbox{\boldmath$\xi$}_I$, also have the 
texture (20).
By comparing Figures (1a) and (1b), one discovers that the difference between
${\mbox{\boldmath$\lambda$}}_I$ and $\mbox{\boldmath$\xi$}_I$ originates from 
the difference between the
supersymmetric interactions of $h$ and the trilinear scalar interactions of $h$.
After the superfield phase redefinitions of (22)
$$
{\mbox{\boldmath$\xi$}}_I = \pmatrix{ 0&-A_4D&0\cr
                          A_4D&0&A_3C\cr
                          0&A_2B & A_1A}_I\eqno(27)
$$
where, in general $A_{1...4}$ are four complex parameters. This pattern, like
that of 
${\bf{ m}}^2_i$, does not lead to flavor-changing difficulties. 
If $A_i$ are all real, then the theory still possesses just two phases, $\alpha$
and $\beta$. If $A_i$ are universal then the $\mbox{\boldmath$\xi$}_I$ and
${\mbox{\boldmath$\lambda$}}_I$
are simultaneously diagonalized.

The CKM matrix is given by
$$
{\bf{ V}} = {\bf{W}}^\dagger_U {\bf{P}}_U {\bf{P}}_D^*  {\bf{W}}_D\eqno(28a)
$$
or
$$
{\bf V} = \pmatrix{1& s_1-s_2e^{i\phi} & s_2s_3\cr
              s_2-s_1e^{i\phi}&e^{i\phi}&-s_3\cr
              -s_1s_3 & s_3 & e^{-i\phi}}\eqno(28b)
$$
where further phase redefinitions have been performed to go from (28a) to (28b)
and
$$
\phi = \alpha + \beta = (\alpha_U - \alpha_D) + (\beta_U - \beta_D),\eqno(29a)
$$
$$
s_1 = s_{12_D}\eqno(29b)
$$
$$
s_2=s_{12_U}\eqno(29c)
$$
$$
s_3=|s_{23_D}e^{i\beta} - s_{23_U}|.\eqno(29d)
$$
The angles $\theta_{1,2,3}$ can all be taken in the first quadrant.
The CP invariant $J$ is given by 
$$
J= Im{V_{ud} V_{tb} V_{ub}^* V_{td}^*} = s_1 s_2 s_3^2 s_\phi.\eqno(30)
$$
Assuming that the observed CP violation in K decays is described by the
standard model box diagrams, 
the measurement of Re $\epsilon$ in CP violation in semileptonic K meson 
decays implies that $s_\phi > 0$, so that $\phi$ is in the first or second
quadrant, depending on the sign of $c_\phi$ which is determined from $|V_{us}|$.
The form (28b) for ${\bf V}$ has been obtained in another context \cite{DHR}
and its consequences explored elsewhere \cite{DHRK,HR}. We stress that, in the
present theory, it is a consequence of a symmetry: the $U(2)$ flavor group.

After superfield rotations to diagonalize the fermion masses, and phase
rotations on scalars to make the neutralino vertices real, as in (24),
the charged wino interactions are
$$
\eqalignno{
[\widetilde{u}^\dagger ({\bf{P}}_U{\bf{P}}_D^*{\bf{W}}_D)d &+
\widetilde{\nu}^\dagger{\bf{W}}_E e]\widetilde{\omega}^+\cr
+[\widetilde{d}^\dagger ({\bf{P}}_D{\bf{P}}_U^*{\bf{W}}_U)u &+
\widetilde{e}^\dagger{\bf{W}}_E\nu]\widetilde{\omega}^- .&(31)\cr}
$$

The $U(2)$ symmetry alone has solved the flavor-changing problem, and produced a
significant economy of parameters in the flavor sector, allowing many
predictions.
Any supersymmetric extension of the standard model must 
involve\footnote{We omit the trilinear parameters in this discussion.}

$\bullet$ 9 quark and lepton masses.

$\bullet$ 15 squark and slepton masses.

$\bullet$ 1 quark mixing matrix, ${\bf{V}}$

$\bullet$ 6 neutralino mixing matrices, ${\bf{W}}_I$ and ${\bf{W}}^c_I$.
The 4 chargino mixing matrices are not independent: ${\bf{W}}^+_q 
={\bf{W}}_U{\bf{V}},
{\bf{W}}^-_q = {\bf{W}}_D {\bf{V}}^\dagger$ and ${\bf{W}}^\pm_e =
{\bf{W}}_E$.

While the hierarchical breaking of $U(2)$ by $\epsilon ' \ll \epsilon \ll 1$
provides an origin for the hierarchy between the fermion masses of the three
generations, the 9 quark and lepton masses remain free parameters. On the other
hand there are only 10 independent squark and slepton masses, since $U(2)$ 
forces $m^2_{2i} = m^2_{1i}$.
The economical achievements of $U(2)$ are mainly in the mixing matrices,
however, and we discuss this below by considering the number of parameters which
enter the quark and lepton masses, and all the mixing matrices.

The lepton sector involves just 4 parameters, $(A,B,C,D)_E$, because the four
phases $(\phi_{A,B,C,D})_E$ can be eliminated.
Once tan $\beta$ is known, three combinations of these $(A, BC/A, AD^2/BC)_E$
are determined by $(m_\tau, m_\mu, m_e)$, leaving just one free
parameter $r_E=(C/B)_E$ for the 4 leptonic gaugino mixing matrices.

In the quark sector there are 10 free parameters: $(A,B,C,D)_{U,D}, \alpha$ and
$\beta$.
The quark masses and CKM matrix involve precisely 10 independent observables, 
so one might guess that these could be used to determine the free parameters.
However, this is not correct.
The quark masses do determine 6 linear combinations of the free parameters:
$(A, BC/A, AD^2/BC)_{U,D}$, leaving four free parameters: $r_{U,D} = 
(C/B)_{U,D}, \alpha$ and $\beta$.
The CKM matrix, ${\bf{V}}$, is parameterized by
$s_1, s_2, s_3, \phi$ of (29). Of these,
$s_1 = \sqrt{m_d/m_s}$ and $s_2 = \sqrt{m_u/m_c}$
depend only on the same combinations of parameters that are determined by the
quark masses.
The parameters $\phi = \alpha + \beta$ and $s_3 = s_3 (r_U, r_D, \beta)$
are determined from $V_{us}$ and $V_{cb}$, and
depend on two combinations of $(r_U, r_D, \alpha, \beta)$. Hence, the
quark masses and ${\bf{V}}$ depend on only 8 of the original 10 parameters.
The two predictions in ${\bf{V}}$ are
$$
\abs{ {V_{td}\over V_{ts}}} = s_1 = \sqrt{ {m_d\over m_s}} = 0.230 \pm 0.
008\eqno(32a)
$$
$$
\abs{ {V_{ub}\over V_{cb}}} = s_2 = \sqrt{ {m_u\over m_c}} = 0.063 \pm 0.
009\eqno(32b)
$$
to be compared with the experimental values of $0.2 \pm 0.1$ and $0.08 \pm 0.
02$ respectively.

The 4 neutralino matrices ${\bf{W}}_{U,D}$ and ${\bf{W}}^c_{U,D}$, of
(24), depend only on the two free parameters $r_{U,D}$, which enter the angles
as shown in (26).
Similarly the two quark chargino mixing matrices, ${\bf{W}}^\pm_q$, shown in
(31), depend only on $r_{U,D}, \alpha$ and $\beta$.

Hence we can summarize the achievements made possible by the introduction of
$U(2)$ and its minimal breaking.

$\bullet$ The supersymmetric flavor-changing problem is solved
and the Yukawa matrices are forced to have a simple
texture, leading to the predictions (32).

$\bullet$ Two small parameters, $\epsilon$ and $\epsilon'$, describe both
the hierarchy of intergenerational fermion masses, and the smallness of
flavor-changing effects induced by superpartner exchange; a structure summarized
by (16) and (17).

$\bullet$ Any supersymmetric extension of the standard model necessarily 
involves
6 new independent flavor mixing matrices, which can be taken as those appearing
at neutral gaugino vertices, ${\bf W}^{(c)}_I$.
In the $U(2)$ theory described above, these 6 new matrices depend on only three
free parameters, $r_I$.

While these results are considerable, the limits to the achievements of $U(2)$
are also apparent. 
There are free parameters for each fermion mass, $V_{cb}$ and for $s_{23_I}$.

The standard model has 12 flavor observables, ignoring CP violation.
Of these, the hierarchy $m_u: m_c: m_t$ can be understood as
$1: \epsilon^2: \epsilon'^2 / \epsilon^2$, and 2
parameters of the CKM matrix are predicted, leaving 7 observables for which 
$U(2)$ provides no understanding.
These 7 remaining pieces of the flavor puzzle can be described in terms of the 
parameters $(A,B,C,D)_I$, defined by the Yukawa matrices in (21):
$$
\eqalignno{
{m_b \over m_\tau} \approx 1  &\Rightarrow {A_D \over A_E} \approx 1 &(33a)\cr
{m_s\over m_b} \approx {1\over 3} {m_\mu\over m_\tau} \approx {1\over 50} 
&\Rightarrow {B_D C_D\over A^2_D} \approx {1\over 3} {B_EC_E\over A^2_E} 
\approx {1\over 50}&(33b)\cr
{m_e m_\mu\over m_\tau^2} \approx {m_dm_s\over m_b^2}
&\Rightarrow {D_D\over A_D} \approx {D_E\over A_E}&(33c)\cr
{m_t \over m_b} \gg 1 &\Rightarrow {A_U v_2 \over A_Dv_1}  \gg 1 &(33d)\cr
{m_c\over m_t} \approx {1\over 10} {m_s\over m_b} &\Rightarrow 
{B_UC_U\over A^2_U} \approx {1\over 10} {B_DC_D\over A^2_D}&(33e)\cr
{m_um_c\over m^2_t} \approx 5.10^{-4} {m_dm_s\over m_b^2} &\Rightarrow
{D_U\over A_U} \approx {1\over 50} {D_D\over A_D}&(33f)\cr
V_{cb} \approx {1\over 25} &\Rightarrow \abs{ {C_D\over A_D} e^{i\beta} - 
{C_U\over A_U} } \approx {1\over 25} &(33g)\cr}
$$
where the approximate equalities hold to better than a factor of 2, and all
parameters and masses are renormalized at the high flavor scale, $M_f$.
A comparison of (33b) and (33g) shows that $B_D \gg C_D$.

As an example, the mass matrices may be
given, at the factor of 2 level and ignoring phases, by
$$
{\bf{m}}_U = 
\pmatrix{ 0&10^{-4}&0\cr
-10^{-4}&0&{x\over 30}\cr
0&{1\over 30x}&1} 175 GeV\eqno(34a)
$$
$$
{\bf{m}}_D = 
\pmatrix{ 
0 & 10^{-3} &0\cr
-10^{-3} & 0 & {y\over 150}\cr
0& {1\over 30y}& {1\over 10} } 25 GeV\eqno(34b)
$$
$$
{\bf{m}}_E = 
\pmatrix{ 0 &10^{-3}&0\cr
-10^{-3}& 0 & {z\over 50}\cr
0 & {1\over 30z } & {1\over 10} } 25 GeV.\eqno(34c)
$$

In section 5 we study the consequences of a $U(2)$ flavor 
symmetry in an $SU(5)$ grand unified theory.
Is such a unified extension possible? If so, can the $SU(5)$ unification shed 
light on any of the patterns and hierarchies of (33) and (34)? Before
addressing these questions, in the next section we extend the analysis for
fermion masses and mixing matrices in the minimal $U(2)$ model to the case that
the rotations in the $23$ sector are large.

\vskip 9pt

\noindent{\bf 4. Large 23 Mixing.}

\vskip 9pt

In $U(2)$ theories, with the minimal texture given in (20), the 23 mixing 
angle in the 
right-handed down sector, $s^c_{23D}$, is expected to be large.
This follows from the observation that $V_{cb}$ and $m_s/m_b$ are of comparable 
magnitude. More precisely, if we forbid $V_{cb}$ from resulting from a 
cancellation of 
large terms in (33g), then $C_D/A_D \ltap 1/10$.
From $m_s/m_b$ of (33b) we deduce that $B_D/D_D \gtap 1/5$.
Thus this 23 mixing in the right-handed down sector is expected to be larger 
than Cabibbo mixing.
A naive estimate gives $s^c_{23D} \approx (m_s/m_b)/V_{cb} \approx 0.5$.
In both $SU(3) \times SU(2) \times U(1)$ and $SU(5)$ theories discussed in this
paper, there are acceptable fits to the data with  $s^c_{23D} \approx 0.3$ so
that the small angle approximation of the previous sector is not a bad first
approximation. However, in both theories there are also good fits to the data
with  $s^c_{23D} \approx 0.7$, which can only be discovered with the analysis 
of this section
.
In this section we derive expressions for mass eigenvalues and mixing matrices 
which treat the $\theta^c_{23}$ diagonalization exactly, while still using 
small angle approximations for $\theta_{23}, \theta_{12}$ and $\theta^c_{12}$.
Rotations in the 23 space yield:\footnote{This analysis applies to $I=U,D$ or
$E$, but for clarity the subscript $I$ is dropped.}
$$
R^T_{23} (s_{23}) \pmatrix{ 0&-D&0\cr
                            D&0&C\cr
                            0&B&A} R^c_{23}(s^c_{23}) =
    \pmatrix{ 0&-Dc^c_{23}& -Ds^c_{23}\cr
              D&-{BC\over A\xi}&0\cr
              0&0&A\xi} \eqno(35)
$$
where $\xi = \sqrt{1+y^2}$ and $y=B/A$ is not necessarily small.
The right-handed mixing angle has
$$
s^c_{23} = {y\over \xi} \ \ \ \ c^c_{23} = {1\over \xi}\eqno(36)
$$
while the left-handed mixing angle is
$$
s_{23} = {1\over y} {m_2\over m_3}\eqno(37)
$$
which is comparable to $m_2/m_3$ for $y$ near unity.
The only small parameter of the heavy 2 $\times$ 2 sector of the Yukawa matrix is
$C/A$, and both $s_{23} = C/\xi^2A$ and $m_2/m_3 = yC/\xi^2A$ are linear 
in $C/A$.
The product
$$
s_{23}s^c_{23} = {1\over \xi} {m_2\over m_3}\eqno(38)
$$
which plays an important role in flavor changing phenomenology,
is reduced by $1/\xi$ compared to the small angle result. 
In the limit that $y$ is small and $\xi = 1 + y^2\rightarrow 1$,
these formulae reduce to the small angle versions of the previous section.
However, even if $y = 1/3$, the $y^2$ correction terms must be kept if 
predictions,
for example for $V_{ub}/V_{cb}$, are to be accurate at the 10\% level.

The right-hand side of (35) shows that the large $\theta^c_{23}$ rotation 
has had two further important consequences:
a non-negligible 13 entry has been generated, requiring an additional
rotation,  $R_{13}$, and the 21 and 12
entries are no longer equal in magnitude, implying that $\theta_{12}$ and 
$\theta^c_{12}$ will have differing magnitudes.
The required diagonalization now has the form
$$
R^T_{13} R^T_{12} R^T_{23}
                        \pmatrix{0&-D&0\cr
                                 D&0&C\cr
                                 0&B&A }  R^c_{23}R^c_{12} = 
                        \pmatrix{ {-AD^2\over BC}&0&0\cr
                                 0&{-BC\over A\xi}&0\cr
                                 0&0&A\xi }           \eqno(39)
$$
where $R_{13}$ is defined with opposite sign to the other rotations
$$
R_{13} = \pmatrix{ 1&0&-s_{13}\cr
                   0&1&0\cr
                   s_{13}&0&1}\eqno(40)
$$
so that $s_{13}$, like $s_{23}, s^c_{23}$ and $s_{12}$, is positive.
We choose all angles to be in the first quadrant, 
except $s^c_{12}$ which is in the second. We find
$$
s_{12}= {1\over \xi^{1/2}} \sqrt{ {m_1\over m_2}} \ \ \ \ \ \ \  s^c_{12} 
= - \xi^{1/2}
\sqrt{ {m_1\over m_2}}\eqno(41)
$$
and
$$
s_{13} = y^2 s_{12}s_{23} = {y\over \xi^{1/2}}
 \sqrt{ {m_1m_2\over m_3^2}}.\eqno(42)
$$
This last result shows that the 13 mixing in the up sector is irrelevant even 
if $y_U$ is of order unity. Such 13 rotations, however, are likely to be 
important for down and lepton sectors.

The matrix ${\bf{W}}^c$ maintains the same form as (24),
except that since $\theta^c_{23}$ is now large, $c^c_{23}$ cannot 
be put to unity:
$$
{\bf{W}}^c = \pmatrix{ 1&s^c_{12}&0\cr
                       -c^c_{23}s^c_{12}&c^c_{23}&s^c_{23}\cr
                       s^c_{12}s^c_{23}&-s^c_{23}&c^c_{23}}\eqno(43)
$$
The matrix ${\bf{W}}$ has a form modified by $R_{13}$
$$
{\bf{W}}= {\bf{R}}_{23} {\bf{R}}_{12}{\bf{R}}_{13} =
\pmatrix{ 1&s_{12}&-s_{13}\cr
          -s_{12}&1&s_{23}\cr
          s_{12}s_{23}+s_{13}&-s_{23}&1}\eqno(44)
$$
so that the $W_{13}$ entry no longer vanishes.   
These neutralinos mixing matrices still conserve CP, and are again
 predicted in terms of just
one free parameter in each of the $U,D,E$ sectors.

The CKM matrix is given by ${\bf{V}}={\bf{W}}^\dagger_U {\bf{P}}_U
{\bf{P}}_D^*{\bf{W}}_D$.
Since $s_{13U}$ is negligible, ${\bf{W}}_U$ is given by (24).
However, $s_{13D}$ is not negligible, so that ${\bf{W}}_D$ has the form of 
(44), hence
$$
{\bf{V}} = 
\pmatrix{ 1&s_1-s_2e^{i\phi}& s_2s_3 - s_{13D} e^{-i(\alpha + \gamma)}\cr
                    s_2-s_1e^{i\phi}&e^{i\phi}&-s_3\cr
                    -s_1s_3+s_{13D}e^{i(\gamma - \beta)} &s_3&e^{-i\phi}
                    }\eqno(45)
$$
where $s_{13D}$ is given by evaluating (42) in the down sector, and the 
phase $\gamma$ is not a new independent phase, but is given by
$$
s_3e^{i\gamma} = s_{23U} - s_{23D} e^{i\beta}
\eqno(46)
$$ 
and cannot be removed from ${\bf{V}}$ when the $0(y^2)$ corrections are kept.
As before, $\phi = \alpha + \beta$, and $\alpha $ and $\beta$ are the two 
physical combinations of phases of the original Yukawa matrices, defined in (22).
It is important to recall that while $s_2 = \sqrt{m_u/m_c}$, and $s_{23U} = 
\sqrt{r_U m_c/m_t}$, the definitions of the angles in the down sector have 
now changed:
$$
\eqalignno{
s_1 &= {1\over \xi^{1/2}} \sqrt{ {m_d\over m_s}}&(47a)\cr
s_{23D} &= {1\over y} {m_s\over m_b}&(47b)\cr
s_{13D} &= {y\over \xi^{1/2}} \sqrt{ {m_dm_s\over m_b^2}}.&(47c)\cr}
$$ 
Treating $\beta$ and $\phi$ as the two independent phases, the predictions for
$\abs{ V_{ub}/V_{cb} }$ and $\abs{ V_{td}/V_{ts} }$ take the form;
$$
\eqalignno{
\abs{ {V_{ub}\over V_{cb} }} &= s_2 
\abs{ 1 + y^2 {s_1\over s_2}
 {s_{23D} \over s^2_3} e^{-i\phi}   (s_{23D} - s_{23U} e^{i\beta}) }&(48)\cr
\abs{ {V_{td}\over V_{ts}}} &= s_1 \abs{ 1 + y^2 {s_{23D} \over s^2_3}
(s_{23D} - s_{23U} e^{-i\beta})}&(49)\cr}
$$
which manifestly display the $O(y^2)$ corrections to the small angle results.
The CP invariant is given by
$$
J = s_1 s_2 s_3^2 s_\phi + y^2 s_1 s_{23D} (s_2 s_{23D} s_\phi + s_2 s_{23U}
s_{\beta-\phi} - s_1 s_{23U} s_\beta). \eqno(50)
$$

It is useful to take the independent phases as $\phi$ and $\beta$, because
$c_\beta$ is determined to be positive by $V_{cb}$, and $c_\phi$ is determined 
from $V_{us}$.
Furthermore, if the $y^2$ correction of (50) does not overwhelm the
$s_1s_2s_3^2s_\phi$ term, then Re$\epsilon$ determines $s_\phi$ to be positive.
In this case the only quadrant ambiguity of the theory is the sign of $s_\beta$.

\vskip 9pt

\noindent{\bf 5. The Minimal $SU(5) \times U(2)$ Model.}

\vskip 9pt

A $U(2)$ flavor symmetry leads to an economical theory
of flavor with Yukawa matrices constrained to have a definite
texture, and neutralino mixing matrices determined in terms
of just three free parameters.
Grand unification provides vertical symmetry relations between the $U,D$ and 
$E$ 
sectors, reducing further the number of flavor parameters.
In this section we study whether the simplest $U(2)$ flavor structure is 
consistent with $SU(5)$ grand unification, and whether the combination of 
these symmetries provides further progress in understanding the pattern of 
quark and lepton masses.

The minimal $SU(5) \times U(2)$ theory is obtained by arranging the light and
heavy matter multiplets into  $10 + \overline{5}$ representations: 
$\chi^a = (T^a, \overline{F}^a), \psi_3 = (t_3, \overline{f}_3)$
and $\psi_a = (t_a, \overline{f}_a)$,
and explicitly writing all $SU(5)$ invariant interactions of 18:
$$
\eqalignno{
\overline{T}_a(M_T T^a + \phi^a t_3 + A^{ab} t_b) &+ 
F_a(M_F\overline{F}^a + \phi^a \overline{f}_3 + A^{ab}\overline{f}_b)\cr
+ h(t_3t_3 + T^at_a) &+ \overline{h}(t_3\overline{f}_3 + T^a\overline{f}_a + 
\overline{F}^at_a)&(51)\cr}
$$
where $h$ and $\overline{h}$ are $5$ and $\overline{5}$ Higgs multiplets.
On integrating out the heavy $T^a, \overline{F}^a$ states, there are 8 
contributions to the Yukawa matrices, shown diagrammatically in Figure 3.

Experiment requires that the Yukawa matrices contain
significant $SU(5)$ breaking at the grand 
unification scale, $M_G$. How can such $SU(5)$ breaking arise?
There are three choices for the insertion of $SU(5)$ breaking: $\phi^a$
or $A^{ab}$ can be $SU(5)$ non-singlets, $T^a$ and $\overline{F}^a$ masses can
contain $SU(5)$ breaking, or additional heavy states can be introduced. 

We prefer to work in the minimal theory described by (51), with $\phi^a$ and
$A^{ab}$ transforming as $SU(5)$ singlets, but with heavy masses:
$$
\eqalignno{
M_T &= M_{T_0} (1 + \epsilon_T Y)&(52a)\cr
M_F &= M_{F_0} (1 + \epsilon_F Y)&(52b)\cr}
$$
where $M_{T_0}$ and $M_{F_0}$ are $SU(5)$ invariant masses, 
which we take to be of order the unification scale, $M_G$.
The $SU(5)$ breaking masses $\epsilon_T M_{T_0}Y$ 
and $\epsilon_F M_{F_0}Y$ arise from the vev of a 24-plet, and are 
proportional to the hypercharge generator, $Y$.
The theory therefore has the tree-level $SU(5)$ breaking of the Yukawa coupling 
matrices isolated in just two parameters, $\epsilon_T$ and $\epsilon_F$.

The Yukawa interactions generated from the 8 diagrams of Figure 8 are
$q^T{\mbox{\boldmath$\lambda$}}_U u^c + q^T {\mbox{\boldmath$\lambda$}}_D 
d^c + e^{cT} {\mbox{\boldmath$\lambda$}}_E \ell$, with 
$$
\eqalignno{
{\mbox{\boldmath$\lambda$}}_U &= \pmatrix{0&-\lambda_4d_U\epsilon'&0\cr
                           \lambda_4 d_U\epsilon' & 0 & \lambda_3 c_U\epsilon\cr
                            0& \lambda_3b_U\epsilon&\lambda_1}&(53)\cr
{\mbox{\boldmath$\lambda$}}_{D,E} &= \pmatrix{ 0 &-\lambda_8d_{D,E}
\epsilon'&0\cr
                        \lambda_8 d_{D,E} \epsilon'& 0& \lambda_5 r c_{D,E} 
                        \epsilon \cr
                        0&\lambda_7b_{D,E}\epsilon& \lambda_2}&(54)\cr}
$$
where $\epsilon = \vev{\phi^2}/M_{T_0}, \epsilon' = \vev{A^{12}}/M_{T_0}$
and $\lambda_1 ... \lambda_8$ are the dimensionless products of trilinear Yukawa
interactions which appear in the diagrams i) ... viii)
of Figure 3, respectively.
The parameter $r = M_{T_0}/M_{F_0}$, while the $SU(5)$ breaking effects 
from the $T^a, \overline{F}^a$
masses are given by the coefficients
$$
b_U = b_D = {1\over 1+{1\over 6} \epsilon_T} \ \ \ \ \ 
b_E = {1\over 1 + \epsilon_T}
\eqno(55a)
$$
$$
c_U = {1\over 1 - {2\over 3} \epsilon_T} \ \ \ \ \ 
c_D = {1\over 1 + {1\over 3} \epsilon_F} \ \ \ \ \ 
c_E = {1\over 1 - {1\over 2} \epsilon_F}\eqno(55b)
$$
$$
\eqalignno{
d_U &= {-{5\over 6} \epsilon_T\over (1-{2\over 3}\epsilon_T)
(1+{1\over 6} \epsilon_T)}&(55c)\cr
d_D &= {1\over 1+ {1\over 6} \epsilon_T} - {\lambda_6\over 
\lambda_8} r {1\over 1 + {1\over 3} \epsilon_F}&(55d)\cr
d_E &= {1\over 1 + \epsilon_T} - {\lambda_6\over \lambda_8} 
r {1\over 1 - {1\over 2} \epsilon_F}.&(55e)\cr}
$$
The labelling of the $\lambda$ parameters allows easy identification
of the diagrammatic origin.
For example, $\overline{F}^a$ exchange occurs in only diagrams v)
and vi), with $\lambda_5$ contributing to the 23 entries of 
${\mbox{\boldmath$\lambda$}}_{D,E}$
and $\lambda_6$ to the 12 entries of ${\mbox{\boldmath$\lambda$}}_{D,E}$.
These contributions are therefore the only ones proportional to $r$.
A close examination of the diagrams of Figure 3 shows that while
$\lambda_1 ... \lambda_7$ are independent parameters,
$\lambda_8 = \lambda_7 \lambda_4/\lambda_3$.

In section 3 we argued that $U(2)$ alone did not address 7 pieces of the 
fermion mass puzzle, as listed in equation (33).
The structure of (53) and (54) shows that the addition of $SU(5)$ unification 
provides an understanding for 4 of these features:

$\bullet$ (33a) The relation $m_b = m_\tau$ at the
unification scale is a well-known success of supersymmetric $SU(5)$.

$\bullet$ (33c) If all dimensionless parameters are taken
to be or order unity, then $m_em_\mu/m_\tau^2 \approx m_dm_s/m_b^2$

$\bullet$ (33f) The anomalously small up quark mass can be understood if the
$SU(5)$ breaking parameter $\epsilon_T$ is small.
The vanishing of $m_u$ in the $SU(5)$ limit follows because the $TTh$ 
interaction gives 
${\mbox{\boldmath$\lambda$}}_U$ symmetric, while $A^{ab}$ is antisymmetric 
and forces the 12 entry to be antisymmetric.
This combination of $SU(5)$ and $U(2)$ symmetry breakings to understand 
the small value of 
$m_u$ is striking, and we consider it a major achievement of the theory.

$\bullet$ ${m_s\over m_b}\approx V_{cb}$.
For textures with vanishing $\lambda_{D22}$ this requires 
$\lambda_{D32} \gg \lambda_{D23}$ or $B_D \gg C_D$, as can be seen by 
comparing (33b) and (33g).
From (54) we see that the $SU(5)$ model can give such a hierarchy if $r$
is small, that is if $M_{F_0} \gg M_{T_0}$.

We note that there is an interesting self-consistency among the last three 
points: 
in the limits that $\epsilon_T, r \to 0$ the determinantal relation
$m_em_\mu/m_\tau^2 = m_dm_s/m_b^2$ becomes exact.

In the limit of small $\epsilon_T$ and $r$, $\epsilon_T$ need only be kept 
in the 12
and 21 entries of ${\mbox{\boldmath$\lambda$}}_U$ and $r$ only in the 23 
entry of ${\mbox{\boldmath$\lambda$}}_{D,E}$.
The Yukawa matrices can then be written
$$
\eqalignno{
{\mbox{\boldmath$\lambda$}}_{U} &= \pmatrix{0&-{5\over 6}\epsilon_T
\epsilon'&0\cr
                           {5\over 6}\epsilon_T\epsilon'&0&\epsilon\cr
                           0&\epsilon&1}\lambda_1&(56a)\cr
{\mbox{\boldmath$\lambda$}}_{D,E} &= \pmatrix{0&-\epsilon'&0\cr
                        \epsilon'&0&r_{D,E}\epsilon\cr
                        0&\epsilon&{1\over \rho} }\lambda_2\rho&(56b)\cr}
$$
where $\epsilon$ and $\epsilon'$ have been rescaled:
$$
\epsilon = {\lambda_3\over \lambda_1} {\vev{\phi^2}\over M_{T_0}} \ \ \ \ 
\epsilon' = {\lambda_4\over \lambda_1} {\vev{A^{12}}\over M_{T_0}}\eqno(57)
$$
$$
r_D = r{1\over 1+ {1\over 3} \epsilon_F} {\lambda_5\over \lambda_7} \ \ \ \ \
r_E = r {1\over 1- {1\over 2} \epsilon_F} \ {\lambda_5\over \lambda_7}\eqno(58)
$$
and
$$
\rho = {\lambda_1\lambda_7\over \lambda_2\lambda_3}.\eqno(59)
$$
A posteriori, the small $\epsilon_T$ approximation turns out to be good to
about 10\%. Although hereafter the corrections in $\epsilon_T$ are neglected in
the explicit analytic formulae, they are kept, as in equations (53) - (55), for
numerical purposes.
In (51) we have assumed that a single 5 or $\overline{5}$ of Higgs, $h$ and 
$\overline{h}$, couple to matter. 
If these contain components of the light Higgs doublets: $h= c_u h_u + ..., 
\; \overline{h} = c_d h_d + ... $
then $c_u$ and $c_d$ should appear as overall
factors in (56a) and (56b) respectively. However, they can be absorbed into
$\lambda_1$ and $\lambda_2$.

In general all parameters appearing in (56a,b) are complex.
however, as discussed in section 3, this texture has only two physical phases, 
$\alpha$ and $\beta$.
In the $SU(5)$ model, these are given by
$$
\eqalignno{
\alpha &= - arg (\epsilon_T)&(60a)\cr
\beta &= - arg \left({1\over 1 + {1\over 3} \epsilon_F} {\lambda_1\lambda_5
\over \lambda_2\lambda_3}\right) &(60b)\cr}
$$
in a basis where $M_{T_0}$ and $M_{F_0}$ are real.
This shows that CP violation can arise only from the $SU(5)$
breaking masses for $T^a$ and $\overline{F}^a$ or from the $\lambda$
parameters, not, for example from the vevs of $\phi^a$ and $A^{ab}$.
If $\epsilon_{T,F}$ were real, we would have $\alpha = 0$ and just a single 
physical phase $\beta = \phi_{CKM}$.
Numerical fits exclude this possibility \cite{BHR}.
Another simplifying possibility is that CP is violated spontaneously only by 
the vev of the 24-plet
which generates $\epsilon_T$ and $\epsilon_F$,
which therefore have a common phase, while the $\lambda$ parameters are
real.
In this paper we take $\alpha $ and $\beta$ to be arbitrary.

After performing the phase rotations of (21),
we can take all parameters of (56a,b) to be real.
The $SU(3) \times SU(2) \times U(1)$ theory of section 3 had 14 flavor 
parameters: $(A,B,C,D)_I, \alpha$ and $\beta$. The $SU(5)$ theory reduces the
number of parameters to 10: $\lambda_1, \lambda_2, \epsilon, \epsilon',
\epsilon_T, \rho, r_D, r_E, \alpha$ and $\beta$. In terms of the $(A,B,C,D)_I$ 
parameters, $SU(5)$ imposes the 4 relations:
$$
A_D = A_E\eqno(61a)
$$
$$
B_D = B_E \eqno(61b)
$$
$$
B_U = C_U \eqno(61c)
$$
$$
D_D = D_E\eqno(61d)
$$
In the limit of small 23 rotation angles, 11 of the 14 parameters of the 
$SU(3) \times
SU(2) \times U(1)$ model are determined from quark and lepton masses and 
mixing, giving the two predictions of (32),
while the 3 free parameters, $r_I = C_I/B_I$,
enter the neutralino mixing matrices, ${\bf{W}}_I$ and ${\bf{W}}^c_I$.
In the $SU(5)$ theory, (61a) and (61d) lead to two further predictions: for
$m_b/m_\tau$ and $m_em_\mu/m_dm_s$, respectively. The two relations (61b) and
(61c) can be viewed as determining two of the free parameters $r_I$: 
$$
\eqalignno{
{r_D \over r_E} &= {m_s \over m_\mu} &(62a)\cr
r_U &= 1 &(62b)\cr}
$$
respectively, so that the mixing matrices 
${\bf{W}}_I$ and ${\bf{W}}^c_I$ depend on only one free parameter.

If the 23 rotation angles of the $D,E$ sectors is large, so that $y \approx 1$,
then (48) and (49) are not necessarily predictions of the theory. In the
$SU(3) \times SU(2) \times U(1)$ theory these predictions are lost:
$V_{ub}/V_{cb}$ and $V_{td}/V_{ts}$
determine two of the free parameters, so that ${\bf{W}}_I$ and ${\bf{W}}^c_I$ 
depend on only a single free parameter. 
In the SU(5) theory, there is only one free parameter, which
is therefore determined by $V_{ub}/V_{cb}$, since it is better measured than
$V_{td}/V_{ts}$, which is predicted from (49). In this case, the 
${\bf{W}}_I$ and ${\bf{W}}^c_I$ are completely predicted.

The analysis for the 3rd generation is not new: $\lambda_1$ and $\lambda_2$ are
determined by $m_t$ and $m_\tau$, allowing a prediction for $m_b$ in terms of 
$\alpha_s$ and $\tan\beta$.
For the second generation we obtain the relations at the unification scale 
$M_G$:
$$
\eqalignno{
{m_c\over m_t} &= \epsilon^2&(63a)\cr
{m_s\over m_b} &= {\rho^2\epsilon^2r_D\over 1+y^2}&(63b)\cr
{m_\mu\over m_\tau} &= {\rho^2\epsilon^2r_E\over 1+y^2}&(63c)\cr
|V_{cb}| &= \epsilon \abs{ e^{i\beta} {\rho r_D \over 1+y^2} - 1}&(63d)\cr}
$$
where $y = \rho\epsilon$, and the $y^2$ correction terms result from the 
large angle diagonalization 
of the 23 space in the $D$ and $E$ sectors, as given in section 4.

The masses of the light generation fermions are obtained from the 
determinants of the Yukawa
matrices
$$
{m_um_c\over m^2_t} = {25\over 36} \epsilon^2_T\epsilon'^2\eqno(64a)
$$
$$
{m_dm_s\over m^2_b} = {m_em_\mu\over m^2_\tau} = {\rho^2\epsilon'^2\over 
(1+y^2)^{3/2}}
\eqno(64b)
$$
The equations of (63a,b,c) and (64a,b) provide 6 constraints, which
can be viewed as determining all the 
remaining parameters, except $\alpha$ and $\beta$.
The CKM matrix is given in (45).
The phase $\phi = \alpha + \beta$ is determined from $\abs{ V_{us}}$,
while a second combination of $\alpha$ and $\beta$ is determined from
$\abs{V_{ub}/V_{cb} }$ via (48).
The ratio
$\abs{V_{td}/V_{ts} }$, or equivalently $J$, can then be viewed as a prediction.

The hierarchy of quark and lepton masses in this $SU(5)$ theory can be 
understood
to be due to the small parameters $\epsilon, \epsilon', \epsilon_T$ and $r$,
with all Yukawa couplings, and hence the $\lambda$ parameters, of order unity.
The single exception to this is that $\rho$ is large, as demonstrated from the 
following  simple estimates, which ignore renormalization group scalings 
and assume that $y$ is not larger than unity.
To avoid a precise cancellation between terms on the right-hand side of (63d) we
require $\epsilon \rho r_D \; \ltap \; V_{cb}$.
This implies from (63b) that $m_s/m_b \; \ltap \; V^2_{cb}/r_D$,
which is why $r_D$ must be small, which we obtained by making $r$ small.
However from (63a) and (63b), 
$m_s/m_b = r_D \rho^2 m_c/m_t$, which requires that $\rho$ be large.

The most plausible origin for large $\rho$
is a small value for $\lambda_2 = \lambda_b$.
Our inability to understand why $\rho$ is large is nothing other than our 
lack of understanding of the large $m_t/m_b$ ratio.
If we insisted on taking $\lambda_2 \approx \lambda_1$ so that $m_t/m_b$ 
arises from a large value for $\tan\beta$, we would be forced to make
$\rho$ large by taking $\lambda_3$ anomalously small.
It seems much more natural to us that $\rho$ is large because the large
$m_t/m_b$ ratio follows from a large 
$(\lambda_1/\lambda_2)$ ratio.
In this case $\tan\beta$ is moderate.
Furthermore, since $\lambda_2 = \lambda_b \ll 1$,
the renormalization group scalings of the masses and mixing angles from $M_G$ 
to weak scales
need only include contributions from $\alpha_s$ and $\lambda_t$.
The CKM matrix is easily scaled by noting that the following quantities are
1 loop renormalization group invariants:
$V_{us}, V_{cu}, V_{ii}, V_{cb} e^{-I_t}, V_{ub}e^{-I_t}, V_{td} e^{-I_t}, 
V_{ts} e^{-I_t}$ and $ Je^{-2I_t}$;
which all follow from the invariants $s_1, s_2, s_3e^{-I_t}$ and $s_\phi$.
For the masses, important invariants are:
$e^{I_t}m_b/\eta_bm_\tau, e^{-3I_t} m_{u,c}/\eta_{u,c}m_t, e^{-I_t} m_{d,s} 
\eta_b/\eta_dm_b$ where $I_t = \int \lambda^2_t dt/4\pi$
and $\eta_i = m_i(m_i)/m_i(m_t)$
for $i=c,b$, whereas for light quarks, $i=u,d,s$,  
$\eta_i = m_i(1GeV)/m_i(m_t)$.
$I_t$ and $\eta_i$ are plotted in Ref. \cite{ADHRS}.
A possible origin for small $\lambda_2$ is that the Higgs multiplets which 
couple to $\psi_3 \psi_3$ are different from those which couple to
$\psi_a\chi^a$.
Small $\lambda_2$ would result if the Higgs multiplet coupling to 
$t_3\overline{f}_3$
contains only a small contribution of the light doublet $h_d$, while other 
Higgs multiplets contain order unity of the light doublets. This would account
for a large value of $\rho$, but otherwise leave our analysis unchanged.

Above we have described how the 10 free flavor parameters of the $SU(5)$ theory
can be determined from data leading to predictions for the three quantities: 
$m_b$, $m_e m_\mu/
m_d m_s$ and $|V_{td}/V_{ts}|$ (or $J$). An alternative procedure is to perform
a $\chi^2$ fit to see how well the model can account for all the relevant data,
which we take to be: the 9 fermion masses, the 3 real CKM mixing angles,
$\epsilon_K$, $\alpha_s$ and the $B^0 \overline{B}^0$ mixing parameter $x_d$.
The predictions for $\epsilon_K$ ($x_d$) involve the quantities 
$B_K$ ($\sqrt{B} f_B$), which we take as further
observables, ``measured" on the lattice. These 17 observables, and their
measured values \cite{GL,L,RPP} are given in Table 1.

\centerline {\bf Table 1}
\begin{center}
\vskip 20pt
\begin{tabular}{|c|c|c|}
\hline
$m_e$ & $0.511$ MeV  & \cr
$m_\mu$ & $105.7$ MeV & \cr
$m_\tau$ & $1777$ MeV & \cr
$(m_u/m_d)_{1 GeV}$ & $0.553 \pm 0.043$ &\cr
$(m_s/m_d)_{1 GeV}$ & $18.9 \pm 0.8$    &\cr
$(m_s)_{1 GeV}$  & $(175 \pm 55)$ MeV &*\cr
$(m_c)_{m_c}$ & $1.27 \pm 0.05$ GeV & \cr
$(m_b)_{m_b}$ & $4.25 \pm 0.15$ GeV & \cr
$(m_t)_{m_t}$ & $165 \pm 10$ GeV & \cr
\hline
$|V_{us}|$ & $0.221 \pm 0.002$ & \cr
$|V_{cb}|$ & $0.038 \pm 0.004$ &* \cr
$|V_{ub}/V_{cb}|$ & $0.08 \pm 0.02$ &* \cr
$|\epsilon_K|$ & $(2.26 \pm 0.02) 10^{-3}$ & \cr
\hline
$\alpha_s (M_Z)$ & $0.117 \pm 0.006$ &* \cr
$x_d$ & $0.71 \pm 0.07$ &* \cr
$\sqrt{B} f_B$ & $(180 \pm 30) MeV$ &* \cr
$B_K$ & $0.8 \pm 0.2$ &* \cr
\hline
\end{tabular}
\end{center}  

These 17 observables depend on 14 parameters: the 10 free flavor parameters,
the ratio of the two electroweak vevs
$v_2/v_1$, $\alpha_s$, $\sqrt{B} f_B$ and $B_K$, so that the fit has 3
degrees of freedom. Since the uncertainties in the 17 observables are very
different, we fix the well measured ones, those without an asterisk in the final
column, to their central values. In particular, inputing central values for 8
of the 9 fermion masses, for $V_{us}$ and for $\epsilon_K$ allows us to 
express 9 of the flavor
parameters and $v_2/v_1$ in terms of the other free parameters. The 7
observables labelled in Table 1 by an asterisk, are then fit by varying the 
1 remaining independent flavor parameter, which we choose to be  $y$, 
and the parameters $\alpha_s, \sqrt{B} f_B$ and $B_K$.
The analysis includes the large 23 mixing results of section 4, and is
therefore not restricted to small $y$. The renormalization scalings from grand
to weak scales include 1 loop contributions from top and strong coupling
constants. For reasons given earlier, we study the case of moderate $\tan
\beta$, so the scalings induced by $b$ and $\tau$ couplings are negligible.

There are three successful fits in which $J$, and therefore $Re \epsilon$, are
positive, as shown in Table 2. 
In fits 1 and 2, $y \approx 0.3$ so that the
$y^2$ correction terms are about 10\%. For these fits $J$ is dominated by $s_1
s_2 s_3^2 s_\phi$ so that $s_\phi$ is positive, and they are distinguished by
the sign of $s_\beta$. In fit 3, $y \approx 1$ and $J$ is dominated by the last
term of (50), so that $s_\beta$ is determined to be negative. 

For each of these
three fits, Table 2 lists the minimum $\chi^2$ values of the seven observables 
which were not set to their central values, the value of
$\chi^2_{min}$ and the corresponding values for 8 of the flavor parameters.
(We leave out $\lambda_1, \lambda_2$ and $v_2/v_1$, which are determined from
the standard analysis of the third generation.)
Finally, the corresponding values for $V_{td}/V_{ts}$ and $J$ are given.
It is clear that each of the fits is extremely good. The analysis of the
uncertainties associated with these fits will be discussed in a separate paper
\cite{BHR}.

Fits 1 and 2 have small y, and in this limit $\sin \beta$ appears only in the
small $y^2$ correction terms of $V_{ub}/V_{cb}$, $V_{td}/V_{ts}$ and $J$, so
the fits are very similar. While $V_{ub}/V_{cb}$ and $J$ have about a 10\%
dependence on the sign of $\sin \beta$, $V_{td}/V_{ts}$ is much less sensitive,
as can be understood from (49). 

\vskip 10pt
\centerline {\bf Table 2}
\begin{center}
\vskip 20pt
\begin{tabular}{|c|c|c|c|}
\hline
&1& 2&3\cr
\hline
sign ($\sin \phi$)&+ &+&$-$\cr
sign ($\sin\beta$)& $-$&+& $-$\cr
$y= \rho \epsilon$& 0.305 & 0.297 & 1.07\cr
\hline
$\alpha_s(M_Z)$&0.117&0.117&0.117\cr
$|V_{cb}|$&0.038&0.040&0.040\cr
$|V_{ub}/V_{cb}|$&0.090&0.071&0.077\cr
$m_s$ /MeV&169&169&164\cr
$f_B\sqrt{B}\;$ /MeV&173&166&187\cr
$x_d$& 0.730&0.738&0.711\cr
$B_K$& 0.875&0.966&0.855\cr
\hline
$\chi^2_{min}$&0.55&1.65&0.55\cr
\hline
$\phi$&1.373&1.367&-2.008\cr
$\beta$&-0.201&0.211&-1.068\cr
$\epsilon$&0.0345&0.0345&0.0359\cr
$\epsilon' \; /10^{-4}$&4.93&5.04&2.36\cr
$\epsilon_T$&0.172&0.168&0.382\cr
$\rho$&8.84&8.61&29.8\cr
$r_D$&0.208&0.219&0.032\cr
$r_E$&0.659&0.694&0.073\cr
\hline
$|V_{td}/V_{ts}|$&0.270&0.267&0.232\cr
$J \; /10^{-5}$&2.63&2.14&2.79\cr
\hline
\end{tabular}
\end{center}

In the Yukawa couplings of (56), and in much of section 5, the full
$\epsilon_T$ dependence of the Yukawa matrices, given in (55), was approximated
by taking $\epsilon_T$ small and keeping only the $\epsilon_T$ dependence in
the numerator of (55c). The results of the numerical fit, which included
the full $\epsilon_T$ dependence, show that this approximation is not very
precise, especially for fit 3.

\newpage
\noindent{\bf 6. Conclusions.}

\vskip 9pt

A $U(2)$ flavor group, broken by small parameters $\epsilon$ and $\epsilon'$,
can solve the supersymmetric flavor-changing problem and provide an
inter-generational fermion mass hierarchy $1:\epsilon^2:\epsilon'^2/\epsilon^2$
\cite{BDH}. The $U(2)$ symmetry leads to successful predictions for
$V_{ub}/V_{cb}$ and $V_{td}/V_{ts}$, and predicts the 6 flavor mixing matrices
at neutralino vertices, ${\bf{W}}_I$ and ${\bf{W}}^c_I$, in terms of just 3 
free parameters $r_{U,D,E}$.

In this paper we have shown that such a $U(2)$ flavor group can be successfully
imposed on an $SU(5)$ grand unified theory, with the consequences that

$\bullet$ Those small quark and lepton mass hierarchies not understood by
$\epsilon$ and $\epsilon'$, and all 3 small angles of the CKM matrix, can be
understood to arise from features of the $SU(5)$ theory.

$\bullet$ The quark and lepton masses, the CKM matrix, and the 6 neutralino
mixing matrices ${\bf{W}}_I$ and ${\bf{W}}^c_I$, are described in terms of 
just 10 flavor parameters (and the ratio of electroweak vevs $v_2/v_1$).

In addition, the Peccei-Quinn $U(1)$ is a sub group of the $U(2)$ flavor
symmetry, and is broken by $\vev{A^{12}} = \epsilon' M_G \approx 3 \times
10^{12}$ GeV, so that the axions are of relevance for the astrophysical dark
matter problem.

Predictions for the 8 fermion mass ratios at the flavor scale are shown in
Table 3, for the cases where the gauge group is $SU(3) \times SU(2) \times 
U(1)$ and $SU(5)$.
The parameters of Table 3 appear in the Yukawa matrices of equation (15) for 
the $SU(3) \times SU(2) \times U(1)$ theory, and in equation (56) for the
$SU(5)$ theory.

For the $SU(5)$ case the predictions are exact, and follow from (56), whereas
in the $SU(3) \times SU(2) \times U(1)$ case, ``$\approx$" means that ratios of
dimensionless couplings are omitted. The $SU(3) \times SU(2) \times U(1)$
theory provides no understanding for many features of the spectrum, for
example,
for why $m_c/m_t \ll m_s/m_b$ or $m_u m_c/m_t^2 \ll m_dm_s/m_b^2$, and must
therefore contain several small dimensionless ratios of Yukawa couplings.

\vskip 10pt
\centerline {\bf Table 3}
\begin{center}
\vskip 20pt
\begin{tabular}{|c|c|c|}
\hline
&$SU(3) \times SU(2) \times U(1)$&SU(5)\cr
\hline
$m_b/m_t$& $\approx 1$& $\lambda_2 / \lambda_1$\cr
$m_b / m_\tau$& $\approx 1$&1\cr
$m_c/m_t$ & $\approx \epsilon^2$& $\epsilon^2$\cr
$m_s/m_b$ & $\approx \epsilon^2$&$\rho^2 \epsilon^2 r_D$\cr
$m_\mu/m_\tau$ & $\approx \epsilon^2$&$\rho^2 \epsilon^2 r_E$\cr
$m_u m_c/m_t^2$ & $\approx \epsilon'^2$ & $(25/36)\epsilon'^2 \epsilon_T^2$\cr
$m_d m_s/m_b^2$ & $\approx \epsilon'^2$ & $\epsilon'^2 \rho^2$\cr
$m_e m_\mu/m_\tau^2$ & $\approx \epsilon'^2$ & $\epsilon'^2 \rho^2$\cr
\hline
\end{tabular}
\end{center}

On the other hand, the $SU(5)$ theory need contain only one small dimensionless
ratio of Yukawa couplings, $\lambda_2/\lambda_1 \ll 1$ to give $m_b/m_t \ll 1$,
with all other hierarchies understood. The parameter $\rho =
(\lambda_1/\lambda_2)(\lambda_7/\lambda_3)$ is expected to be large (due to the
large $\lambda_1/\lambda_2$ ratio), explaining why $m_s/m_b$ and 
$m_\mu/m_\tau$ are
larger than $m_c/m_t$, and contributing to the understanding of $m_u m_c/m_t^2
\ll m_d m_s/m_b^2, m_em_\mu/m_\tau^2$. The anomalously low value for $m_u
m_c/m_t^2$ is understood in terms of a small amount of $SU(5)$ breaking,
$\epsilon_T$, in the mass of the heavy 10-plet: $M_T = M_{T_0}(1 + \epsilon_T
Y)$.
The vanishing of $m_u$ in the $SU(5)$ symmetric limit is particularly striking:
the $T_a A^{ab} T_b h$ coupling is made antisymmetric by $U(2)$ invariance, but
symmetric by $SU(5)$ invariance. The only $SU(5)$ breaking in the Yukawa
matrices at the unification scale is due to $\epsilon_T \neq 0$ and $r_D \neq
r_E$. Since $m_\mu/m_s = r_E/r_D = (1+\epsilon_F/3)/(1-\epsilon_F/2)$ is close
to 3, the fractional breaking of $SU(5)$ in the mass of the heavy 5-plet,
$\epsilon_F$, is of order unity, where $M_F = M_{F_0}(1 + \epsilon_F Y)$.

The consequences of the $U(2)$ flavor symmetry are similar in the 
$SU(3) \times SU(2) \times U(1)$ and $SU(5)$ theories. In the small 23
rotation angle approximation, valid for fits 1 and 2 of the previous section,
the CKM matrix is parameterized by the 4 angles $s_1, s_2, s_3$ and $s_\phi$.
The parameters $s_1$ and $s_2$ are determined by quark mass ratios $s_1 =
\sqrt{m_d/m_s}$ and $s_2 = \sqrt{m_u/m_c}$, so that the sizes of 
$V_{ub}/V_{cb}$,  $V_{td}/V_{ts}$ and $V_{us}$ are automatically understood in
$U(2)$ theories in terms of quark mass hierarchies. This is not the case for
$s_3$, which also depends on the parameters $r_U, r_D$ and $\beta$:
$s_3 = |\sqrt{r_D m_s/m_b} e^{i \beta} - \sqrt{r_U m_c/m_t}|$. (The only
difference in the expressions for the CKM parameters in the  
$SU(3) \times SU(2) \times U(1)$ and $SU(5)$ theories, is that, as discussed
below, $r_U = 1$ in the $SU(5)$ case.) The observed value of $V_{cb}$ therefore
requires that $r_D$ is small. In the $SU(5)$ theory this can be understood as
arising from $r = M_{T_0}/M_{F_0} < 1$. \footnote{It is perhaps surprising that
$\epsilon_T \ll \epsilon_F$, given that the T is lighter than the F. However,
in practice $r \approx 1/5$, and is not very small.}
Hence, in the $SU(5)$ theory, all small quark and lepton mass ratios, 
and the small values of all three CKM mixing angles, can be understood in terms
of three small symmetry breaking parameters, $\epsilon, \epsilon'$ and
$\epsilon_T$, and the ratio of heavy masses, $r$. The only exception is the
small ratio $\lambda_2/\lambda_1$.

The CP violating phase $\phi$ is determined to have a large magnitude from
$|V_{us}| =|\sqrt{m_d/m_s} - e^{i \phi} \sqrt{m_u/m_c}|$. The size of CP
violation can therefore be determined from CP conserving quantities $-$ quark 
mass ratios and the CKM flavor mixing angles $-$ and is a significant success
of the $U(2)$ symmetry.

In going from the $SU(3) \times SU(2) \times U(1)$ theory to the $SU(5)$
theory, the number of independent flavor parameters is reduced from 14 to 10.
The parameter relations imposed by $SU(5)$ are shown in (61). They directly
give
$$
\eqalignno{
m_b &= m_\tau &(65a)\cr
{r_D \over r_E} &= {m_s \over m_\mu} &(65b)\cr
r_U &= 1 &(65c)\cr
{ m_e m_\mu \over m_\tau^2} &= {m_d m_s \over m_b^2} &(65d)\cr}
$$
at the unification scale.
The success of (65a) is a well-known feature of supersymmetric $SU(5)$. The
$SU(5)$ mass relation (65d) is less well-known, but is equally successful.
Although such a relation has been obtained before \cite{ADHRS}, in the present
theory it is a consequence of a texture forced by the $U(2)$ flavor symmetry. 
The relations (65b) and (65c) reduce the number of free parameters entering the
6 neutralino mixing matrices, ${\bf{W}}_I$ and ${\bf{W}}^c_I$, from 3 to 1:
$$
\eqalignno{
{\bf{W}}_I &=    \pmatrix{1&\sqrt{{m_1 \over m_2}}&0\cr
            -\sqrt{{m_1 \over m_2}} & 1 & \sqrt{r{m_2 \over m_3}}\cr
            \sqrt{r {m_1 \over m_3}} & -\sqrt{r {m_2 \over m_3}} &1}_I &(66a)\cr
{\bf{W}}^c_I &=    \pmatrix{1&-\sqrt{{m_1 \over m_2}}&0\cr
                      \sqrt{{m_1 \over m_2}} & 1 & \sqrt{{1 \over r} {m_2 \over
                      m_3}}\cr
                       \sqrt{{1 \over r} {m_1 \over m_3}} & -\sqrt{{1 \over r}
                       {m_2 \over m_3}} &1}_I &(66b)
                       \cr}
$$ 
with $r_U = 1$ and $r_D/r_E \approx 1/3$. For the case of large 23 rotation 
angles in the $D,E$ sectors, as in fit 3 of section 5, the forms of the CKM and 
${\bf{W}}_I$ and ${\bf{W}}^c_I$ matrices are more complicated. 
While $V_{ub}/V_{cb}$ can no
longer be viewed as a prediction, there are no free parameters at all in  
${\bf{W}}_I$ and ${\bf{W}}^c_I$.  If
supersymmetry is discovered, this theory can be tested by the predictions
(66a,b) for ${\bf{W}}_I$ and ${\bf{W}}^c_I$.

The $U(2)$ theory of flavor presented in this paper makes definite predictions
for various processes, as will be discussed in a separate paper \cite{BHR}.
However, the $U(2)$ symmetry is insufficient to determine the fractional mass
splittings between the scalars of the third generation and the scalars of the
lighter two generations, $\Delta_L$ and $\Delta_R$ for the left and right
components respectively. If $\Delta_L = \Delta_R = 1$ in the down sector,
then, in the $SU(5)$ theory discussed in this paper, the gluino exchange
contribution to $\epsilon_K$ exceeds the experimental value by about a factor
of 50, for average squark masses and a gluino mass of 1 TeV. Hence, in the
down sector of a $U(3)$ theory of flavor, it will be crucial to either suppress
$\Delta_L$ and/or $\Delta_R$, or to have milder flavor mixings to the third 
generation than given by (66a,b).

\noindent{\bf Acknowledgements.}

\vskip 9pt

We are deeply indebted to Andrea Romanino for help with the numerical fits of 
section 5 and for several very useful remarks.

\end{document}